\documentclass{article}%
\usepackage{graphicx}
\usepackage{amsmath}
\usepackage{amsfonts}
\usepackage{amssymb}%
\begin{document}

\title{Quantum dynamics under coherent and incoherent effects of a spin bath in the
Keldysh formalism: application to a spin swapping operation}
\author{Ernesto P. Danieli, Horacio M. Pastawski\thanks{Corresponding author. E-mail
horacio@famaf.unc.edu.ar}, and Gonzalo A. \'{A}lvarez.\\Facultad de Matem\'{a}tica, Astronom\'{\i}a y F\'{\i}sica,\\Universidad Nacional de C\'{o}rdoba,\\Ciudad Universitaria, 5000 C\'{o}rdoba, Argentina.}
\maketitle

\begin{abstract}
We develop the Keldysh formalism for the polarization dynamics of an open spin
system. We apply it to the swapping between two qubit states in a model
describing an NMR cross-polarization experiment. The environment is a set of
interacting spins. For fast fluctuations in the environment, the analytical
solution shows effects missed by the secular approximation of the Quantum
Master Equation for the density matrix: a frequency decrease depending on the
system-environment escape rate and the quantum quadratic short time behavior.
Considering full memory of the bath correlations yields a progressive change
of the swapping frequency.

\end{abstract}

\section{Introduction}

The characterization and control of spin dynamics in open and closed spin
systems of intermediate size remain a problem of great interest
\cite{z--Loss-PRL03}. Recently, such systems have become increasingly
important in the emerging field of quantum information processing
\cite{z--BennettDiVicc}. The quantum interferences of these systems become
damped by the lack of isolation from the environment and one visualizes this
phenomenon as decoherence. Indeed, the inclusion of the degrees of freedom of
the environment may easily become an unsolvable problem and requires
approximations not fully quantified. This motivates a revival of interest on
previous studies in various fields such as Nuclear Magnetic Resonance
\cite{z--MKBE74}, quantum transport \cite{z--DaPa1990} and the
quantum-classical correspondence problem \cite{z--PazZurek,z--Zurek} with a
view on their application to emergent fields like the quantum computation
\cite{z-LidarWu,z--De Raedt,z--Cory} and molecular electronics
\cite{z--MolElec+Deco0,z--MolElec+Deco,z--MolElec1,z--MolElec2}.

The most standard framework adopted to describe the system-environment
interaction is the use of the Quantum Master Equation, derived from the
Liouville-von Newman equation \cite{z--QmasterE1,z--QmasterE2} in a fast
fluctuation approximation. Interactions with the environment occur at a rate
given by the Fermi Golden Rule (FGR) providing a dissipative mechanism that
could induce a non unitary dynamics into the system. An overall (conservation)
balance condition is obtained by imposing a convergence into the thermal
equilibrium state. While sufficient for most traditional applications, this
approximation leaves aside important memory effects and interferences in the
time domain produced by the coherent interaction between the system and the
bath which are becoming of increasing interest \cite{z--Taylor}.

The present work focuses on two spin correlation functions in small open
systems with environmental interactions under conditions where the dynamical
feedback effects, that go beyond the Fermi Golden Rule, become relevant. For
this, we will resort to the Keldysh non-equilibrium formalism which leads to
an integral solution of the Schr\"{o}dinger equation. While this novel
situation is presented here for the first time, the formalism already inspired
original experimental and theoretical developments in coherent spin dynamics
involving quantum interferences in the time domain. In particular, it was used
to develop the notion of polarization waves leading to mesoscopic echoes
\cite{z--EcosMesoscI,z--EcosMesoscII}, to establish the influence of chaos
on\ time reversal (Loschmidt echoes) \cite{z--LUP-JCP1998,z--JalPa} and to
establish the possibility of a spin projection chromatography \cite{z-SPC}. A
rough account of many-body decoherence enabled the interpretation \ of
anomalies in spin \textquotedblleft diffusion\textquotedblright\ experiments
as a manifestation of the quantum Zeno effect \cite{z-QZE}. We now make a leap
forward in the development of this formalism\ by showing how it deals with
open systems. The application to\ a case with an exact analytical solution
\cite{z--Nuestro2002} and where more standard approximations can be obtained
\cite{z--MKBE74} will show the potential of our proposal.

\section{The Keldysh formalism for open systems}

In this section we make a brief introduction to the Keldysh formalism,
summarizing our results for closed systems of Ref. \cite{z-SPC}. Our aim is to
extend them to open systems. Let us start considering a system with $M$ spins
1/2. The spin correlation function,
\begin{equation}
P_{m,n}(t)=\frac{\left\langle \Psi_{eq}\right\vert \widehat{S}_{m}%
^{z}(t)\widehat{S}_{n}^{z}(0)\left\vert \Psi_{eq}\right\rangle }{\left\langle
\Psi_{eq}\right\vert \widehat{S}_{n}^{z}(0)\widehat{S}_{n}^{z}(0)\left\vert
\Psi_{eq}\right\rangle }, \label{eq--PCFunc}%
\end{equation}
gives the amount of the $z$ component of the local polarization at time $t$ on
$m$-th site, provided that the system was, at time $t=0,$ in its equilibrium
state with a spin \textquotedblleft up\textquotedblright\ added at the $n$-th
site. Here, $\widehat{S}_{m}^{z}(t)=e^{\mathrm{i}\mathcal{H}t/\hbar}%
\widehat{S}_{m}^{z}e^{-\mathrm{i}\mathcal{H}t/\hbar}$ is the spin operator in
the Heisenberg representation and $\left\vert \Psi_{eq}\right\rangle =\sum
_{N}a_{N}\left\vert \Psi_{eq}^{(N)}\right\rangle $ is the thermodynamical
many-body equilibrium state constructed by adding states with different number
$N$ of spins up with appropriate statistical weights and random phases. The
Jordan-Wigner transformation\ (JWT)
\cite{z--XY-dynam-theoI,z--XY-dynam-theoII} $\widehat{S}_{n}^{+}=\widehat
{c}_{n}^{+}\exp\{\mathrm{i}\pi\sum_{m=1}^{n-1}\widehat{c}_{m}^{+}\widehat
{c}_{m}^{{}}\},$ establishes the relation between spin and fermion\ operators
at site $n.$ Symbols $\widehat{c}_{n}^{+}$ and $\widehat{c}_{n}^{{}},$ stand
for the fermion creation and destruction operators, and $\widehat{S}_{n}^{\pm
}$ are the rising and lowering spin operator $\widehat{S}_{n}^{\pm}%
=\widehat{S}_{n}^{x}\pm\mathrm{i}\widehat{S}_{n}^{y}$, where $\widehat{S}%
_{n}^{\;u}$ ($u=x,y,z$) represents the Cartesian spin operator. The initial
polarized state is described by the non-equilibrium state $\left\vert
\Psi_{\mathrm{n.e.}}\right\rangle =\widehat{c}_{n}^{+}\left\vert
\Psi_{\mathrm{eq.}}\right\rangle $ formed by creating an excitation in the
$n$-th site at $t=0$. Its further evolution is contained in the particle
density function \cite{z--Keldysh,z--Mahan} in the Keldysh formalism
$G_{m,n}^{<}(t_{2},t_{1})=\tfrac{\mathrm{i}}{\hbar}\left\langle \Psi
_{\mathrm{n.e.}}\right\vert \widehat{c}_{m}^{+}(t_{1})\widehat{c}_{n}^{{}%
}(t_{2})\left\vert \Psi_{\mathrm{n.e.}}\right\rangle $ which can be split into
contributions $G_{m,n}^{<\,_{(N)}}(t_{2},t_{1})$ from each subspace with $N$
particles (or equivalently $N$ spins up). Considering that we are in the high
temperature regime, i.e., $k_{B}T$ \ is much\ higher than any energy scale of
the system, this enables us to re-write Eq. (\ref{eq--PCFunc}) as \cite{z-SPC}%

\begin{align}
P_{m,n}(t) &  =\tfrac{2\hbar}{\mathrm{i}}G_{m,m}^{<\,}%
(t,t)-1,\,\,\mathrm{with}\label{eq--N_PCFunc}\\
G_{m,m}^{<\,}(t,t) &  =\sum_{N=1}^{M}\dfrac{\left(
\genfrac{}{}{0pt}{1}{M-1}{N-1}%
\right)  }{2^{M-1}}G_{m,m}^{<\,_{(N)}}(t,t).\label{eq--GmenosTotal}%
\end{align}
Notice that the non-equilibrium density $G_{m,m}^{<\,}(t,t)$ depends
implicitly on the index $n$ that indicates the site of the initial ($t=0$)
excitation. The expression for this initial condition is%

\begin{equation}
G_{k,l}^{<_{(N)}}(0,0)=\tfrac{\mathrm{i}}{\hbar}\left(  \tfrac{N-1}{M-1}%
\delta_{k,l}+\tfrac{M-N}{M-1}\delta_{k,n}\delta_{n,l}\right)  .
\label{eq--initial-density}%
\end{equation}
Here the first term is the equilibrium density and it can be seen that is
identical for all the sites. The second term represents the non-equilibrium
contribution where only the $n$-th site is different from zero. In general,
this density function evolves under the Schr\"{o}dinger equation expressed in
the Danielewicz form \cite{z--Danielewicz}, which becomes:%

\begin{gather}
G_{m,m}^{\;<_{(N)}}(t_{2},t_{1})=\hbar^{2}\sum_{l,k}G_{m,k}^{R\,_{(N)}}%
(t_{2},0)G_{k,l}^{\;<_{(N)}}(0,0)G_{l,m}^{A_{(N)}}(0,t_{1})\nonumber\\
+\sum_{l,k}\int_{0}^{t_{2}}\int_{0}^{t_{1}}G_{m,k}^{R_{(N)}}(t_{2}%
,t_{k})\Sigma_{k,l}^{\,\,<_{(N)}}(t_{k},t_{l})G_{l,m}^{A_{(N)}}(t_{l}%
,t_{1})\mathrm{d}t_{k}\mathrm{d}t_{l}.\label{eq--Gmenos-Nsubspace}%
\end{gather}
Here $G_{m,k}^{R_{(N)}},$ and $G_{k,m}^{A_{(N)}}$ are the exact retarded
($t_{2}>t_{1}>0$) \ and advanced ($0<t_{2}<t_{1}$) two particle propagators or
Green's functions of the many-body system.

The first term in the rhs of Eq. (\ref{eq--Gmenos-Nsubspace}) can be seen as a
generalization of the integral form of the (reduced) density matrix
($\rho(t)=\mathrm{e}^{-\mathrm{i}\mathcal{H}t/\hbar}\rho(0)\mathrm{e}%
^{\mathrm{i}\mathcal{H}t/\hbar}$) projected over a basis of single particle
excitations in its real space representation. This term is all one needs to
solve systems such as a finite or infinite one dimensional chain with nearest
neighbors $XY$ interaction \cite{z-SPC}. In contrast, systems with topological
defects \cite{z--Nuestro2002}, long range interaction or Ising terms in the
spin Hamiltonian present complex many-body effects in the particle
description. These lead to mean-life, $\mathrm{Im}\Sigma_{{}}^{\,\,R},$ of
the\ single particle states, producing the non-conservation of probability on
the retarded and advanced propagators, $G_{{}}^{R\,}$ and $G_{{}}^{A\,}$. In
this case, the second term would collect incoherent reinjections, given by
$\Sigma_{{}}^{<},$ that compensate any eventual \textquotedblleft
leak\textquotedblright\ from the coherent evolution. They also can account for
processes not conserving the spin projection. A key idea in this paper is that
a similar effect of density non-conservation appears when one attempts to
reduce the whole $XY$ system into a \textquotedblleft system\textquotedblright%
\ of $2$ spins and an \textquotedblleft environment\textquotedblright\ with
$M-2$ spins. Under these conditions the sum in Eq. (\ref{eq--GmenosTotal})
will run only over the subspaces allowed within the \textquotedblleft
system\textquotedblright, $N=1,2.$ The effects of the \textquotedblleft
environment\textquotedblright\ will be included in the form of self-energy
terms, $\Sigma_{k,l}^{\,\,<}(t_{k},t_{l})$ and $\Sigma_{k,l}^{\,\,R}%
(t_{k},t_{l})$ modifying the reduced \textquotedblleft
system\textquotedblright.

If we replace Eq. (\ref{eq--Gmenos-Nsubspace}) into Eq. (\ref{eq--GmenosTotal}%
) and perform the summation in the $N$ index only over the \textquotedblleft
system\textquotedblright, the result can be seen as the sum of two
contributions reproducing the structure of Eq. (\ref{eq--Gmenos-Nsubspace}).
Then, the first term will be called the \textit{coherent} contribution because
it is related to the initial condition within the \textquotedblleft
system\textquotedblright. The evolution of this initial \textit{density}%
\ decays with time $t$ as a consequence of its escape towards the region
called the \textquotedblleft environment\textquotedblright.

The second term will account for the thermodynamical nature of the
\textquotedblleft environment\textquotedblright\ when $M\rightarrow\infty$. It
can be seen as a boundary condition that modifies the density of the
\textquotedblleft system\textquotedblright. If the mean occupation of the
\textquotedblleft environment\textquotedblright\ is lower than that of the
\textquotedblleft system\textquotedblright, there will be a flux of
probability from the \textquotedblleft system\textquotedblright\ to the
\textquotedblleft environment\textquotedblright\ included in the formalism
through the retarded and advanced propagators. On the other hand, if the
\textquotedblleft environment\textquotedblright\ mean occupation were higher
than that of the \textquotedblleft system\textquotedblright, it would
establish a probability flow from the \textquotedblleft
environment\textquotedblright\ to the \textquotedblleft
system\textquotedblright\ and this could be seen as if the \textquotedblleft
environment\textquotedblright\ were injecting probability into the
\textquotedblleft system\textquotedblright. The evolution of this injected
probability is described by the second term in Eq. (\ref{eq--GmenosTotal})
which will be called the \textit{incoherent }contribution. Thus, the
probability within the \textquotedblleft system\textquotedblright\ would be
fed by the \textquotedblleft environment\textquotedblright.

In general terms Eq. (\ref{eq--GmenosTotal}) and (\ref{eq--Gmenos-Nsubspace})
involve two time functions. In order to take profit of the information hidden
in the time correlations, it is convenient to use the new time-energy
variables [$t,\varepsilon$] \cite{z--GLBEII}. This is inspired in the Wigner
coordinates that exploit the spatial correlations to define the
position-momentum variables [$x,p_{x}$]. Appendix A shows how this procedure
is performed. Applying this technique to Eq.(\ref{eq--Gmenos-Nsubspace}) we obtain%

\begin{multline}
G_{m,m}^{\;<\,_{(N)}\,}(t,t)=\label{eq--DanielET}\\
\hbar^{2}\int_{-\infty}^{\infty}\int_{-\infty}^{\infty}\sum_{k,l=-1}%
^{0}G_{m,k}^{R_{(N)}\,}(\varepsilon+\tfrac{\hbar\omega}{2})G_{k,l}%
^{\;<\,_{(N)}}(0,0)G_{l,m}^{A_{(N)}\,}(\varepsilon-\tfrac{\hbar\omega}{2}%
)\exp(-\mathrm{i}\omega t)\frac{\mathrm{d}\omega}{2\pi}\frac{\mathrm{d}%
\varepsilon}{2\pi\hbar}\\
+\int_{0}^{t}\int_{-\infty}^{\infty}\int_{-\infty}^{\infty}\sum_{k,l=-1}%
^{0}G_{m,k}^{R\,_{(N)}\,}(\varepsilon+\tfrac{\hbar\omega}{2})\Sigma
_{k,l}^{\;<\;_{(N)}}(\varepsilon,t_{i})G_{l,m}^{A_{(N)}\,}(\varepsilon
-\tfrac{\hbar\omega}{2})\times\\
\exp\{-\mathrm{i}\omega(t-t_{i})\}\frac{\mathrm{d}\omega}{2\pi}\frac
{\mathrm{d}\varepsilon}{2\pi\hbar}\mathrm{d}t_{i}.
\end{multline}

We will apply this formalism to a system of $M$\textbf{\ }spins$~\frac
{\mathrm{1}}{\mathrm{2}}$ arranged in a chain. Their interaction through an
$XY$ coupling enables the swapping between nearest neighbor spins. In the
presence of a magnetic field, the Hamiltonian is
\begin{equation}
\widehat{\mathcal{H}}^{\mathrm{chain}}=\sum_{n=1}^{M}\hbar\Omega_{n}^{{}%
}\left[  \widehat{S}_{n}^{+}\widehat{S}_{n}^{-}-\tfrac{1}{2}\right]
+\tfrac{1}{2}\sum_{n=1}^{M-1}J_{n,n+1}^{{}}(\widehat{S}_{n}^{+}\widehat
{S}_{n+1}^{-}+\mathrm{c.c.}), \label{eq--HchainI}%
\end{equation}
which has a Zeeman part, $\widehat{\mathcal{H}}_{Z}^{{}}$, proportional to
$\widehat{S}_{n}^{z}$ with $\hbar\Omega_{n}^{{}}$ the Zeeman energy; and a
swapping (flip-flop) term, $\widehat{\mathcal{H}}_{XY}^{{}},$
where\ $J_{n,n+1}^{{}}$ is the coupling between sites $n$\textit{\ }and
$n+1$\textit{. }

This simplified model can be used as an approximation to real $^{13}$C$-^{1}$H
systems in an NMR cross-polarization (CP) experiment
\cite{z--Pines72,z--MKBE74}. We will model the $^{13}$C and $^{1}$H nuclei,
close to the Hartmann-Hahn condition, as the first two sites of the linear
chain and the rest of the chain would represent the proton spin bath or
\textquotedblleft environment\textquotedblright.

Instead of solving a high dimensional spin Hamiltonian (\ref{eq--HchainI})
describing the \textquotedblleft system\textquotedblright\ plus the
\textquotedblleft spin bath\textquotedblright, the JWT provides a map into a
fermionic system. For a one dimensional chain or ring with nearest neighbor
interactions the dimension of the Hilbert space can be reduced from 2$^{M}$ to
$M$ enabling the calculation of different aspects of spin dynamics
\cite{z--EcosMesoscI,z--EcosMesoscII,z--Nuestro2002} and quantum coherences
\cite{z--FeldLacelle}. Since the interaction is restricted to nearest
neighbors, the only non-zero coupling terms are proportional to $\widehat
{c}_{n}^{+}\widehat{c}_{n+1}^{{}}=\widehat{S}_{n}^{+}\widehat{S}_{n+1}^{-}$.
Each subspace with $({%
\genfrac{}{}{0pt}{}{M}{N}%
})$ states of spin projection $\left\langle \sum_{n=1}^{M}\widehat{S}_{n}%
^{z}\right\rangle =N-M/2$ is now a subspace with $N$ \textit{non-interacting}
fermions. The eigenfunctions $\left\vert \Psi_{\gamma}^{(N)}\right\rangle $ of
these sub-spaces are expressed as \textit{single} Slater determinants built-up
with the single particle wave functions $\varphi_{\alpha}$ of energy
$\varepsilon_{\alpha}$. Under these conditions $G_{m,n}^{R\,_{(N)}}%
=G_{m,n}^{R\,}$\ for all $N.$

\section{\textbf{A two-spin system connected to a spin bath}}

We label the \textquotedblleft system\textquotedblright\ sites with the
numbers $-1$ for the $^{13}$C and $0$ for the $^{1}$H containing the initial
excitation. Thus, we want to obtain an analytical expression for the local
polarization at each site of the \textquotedblleft system\textquotedblright%
\ that, according to Eq. (\ref{eq--N_PCFunc}), is proportional to the particle
density Green's function $G_{i,i}^{<}(t)$ (for $i=-1,0$).

In these conditions, the effective (reduced) Hamiltonian is:
\begin{equation}
\mathbf{\tilde{H}}=\left(
\begin{array}
[c]{cc}%
E_{-1} & V\\
V & E_{0}+\Sigma_{0}^{R}%
\end{array}
\right)  ,\label{eq--Ham2sites}%
\end{equation}
where $E_{-1}\equiv\hbar\Omega_{-1}^{{}}$ and $E_{0}\equiv\hbar\Omega_{0}^{{}%
}$, are site energies and $V=\frac{J}{2}$ represents the swapping strength.
For simplicity we will take $E_{-1}=E_{0}$. The self-energy $\Sigma_{0}^{R}$
renormalizes the site energy of the $0$-th site due to the action of the rest
of the chain \cite{z--LevPasDAmato}. This procedure makes possible to trace
out all the degrees of freedom of the environment without loss of information.
It satisfies
\[
\Sigma_{0}^{R}(\varepsilon)=(\frac{V_{0}}{V_{c}})^{2}\Sigma_{1}^{R}%
(\varepsilon),
\]
where $V_{0}$ is the system-environment coupling through the $0$-th site. In a
finite chain $\Sigma_{1}^{R}(\varepsilon)$ can be calculated using the
recurrence relations
\begin{equation}
\Sigma_{i}^{R}(\varepsilon)=\frac{V_{c}^{2}}{\varepsilon-E_{c}-\Sigma
_{i+1}^{R}(\varepsilon)},\label{eq--SelfEnergy}%
\end{equation}
stating with $\Sigma_{M}^{R}(\varepsilon)=0.$ Here $E_{c}$ is the center of
the energy band of the homogeneous linear chain that is acting as the
environment, $V_{c}$ is the nearest neighbor hopping within the chain. In this
case, $\Sigma_{0}^{R}(\varepsilon)$ is the ratio between polynomials of degree
$M-3$ and $M-2$ on $\varepsilon.$ The roots of the denominator are the $M-2$
eigenenergies of the \ environment. This functional dependence accounts
exactly for the memory effects in the \textquotedblleft
environment\textquotedblright\ and describes a variety of interference
phenomena such as quantum beats and mesoscopic echoes. In order to include the
effect of irreversible loss of information and simplify the calculations, we
let the number of spins conforming the linear chain tends to infinity, that
is, $M\rightarrow\infty.$ On that situation, $\Sigma_{i}^{R}(\varepsilon
)=\Sigma_{i+1}^{R}(\varepsilon)=\Sigma^{R}$ and Eq. (\ref{eq--SelfEnergy})
becomes a Dyson equation \cite{z--DaPa1990}. When the energy $\varepsilon$
lies within the band of propagating excitations, $\left\vert \varepsilon
-E_{c}\right\vert \leq2\left\vert V_{c}\right\vert ,$ the solution is
\begin{align}
\Sigma^{R}(\varepsilon) &  =\Delta_{c}(\varepsilon)-\mathrm{i}\Gamma
_{c}(\varepsilon)\label{eq--exact-Self-Energy}\\
&  =\frac{\varepsilon-E_{c}}{2}-\mathrm{i}\sqrt{V_{c}^{2}-(\frac
{\varepsilon-E_{c}}{2})^{2}}.\nonumber
\end{align}
For the present problem we will work with the assumption that $\left\vert
E_{0}-E_{-1}\right\vert ,V,V_{0}\ll V_{c}.$ This means that the temporal
fluctuations of the environment are faster than any characteristic time of the
adopted model. This approximation allows us to consider that
$\operatorname{Im}\Sigma_{0}^{R}(\varepsilon)\simeq(\frac{V_{0}}{V_{c}}%
)^{2}\Gamma_{c}(E_{0})=\Gamma$ and the level $E_{0}$ becomes broaden according
to the Fermi Golden Rule. Since typically, $E_{0}\simeq E_{c},$ the
corresponding shift $\operatorname{Re}\Sigma_{0}^{R}(\varepsilon)\simeq
(\frac{V_{0}}{V_{c}})^{2}\Delta_{c}(E_{0})$ is a small correction that can be neglected.

In order to obtain the contribution of the coherent term in Eq.
(\ref{eq--DanielET}), we need to compute the Fourier transform of the product
of two propagators $G_{m,k}^{R\,}(\varepsilon+\tfrac{1}{2}\hbar\omega
)G_{l,m}^{A\,}(\varepsilon-\tfrac{1}{2}\hbar\omega)$ obtained as matrix
elements of the resolvent $\mathbf{G}(\varepsilon)=\left\vert \varepsilon
\mathbf{I-\tilde{H}}\right\vert ^{-1}.$ One then integrates over the energy
variable $\varepsilon$.

\begin{figure}[ptbh]
\centering
\includegraphics[trim=7cm 7cm 7cm 7cm,natheight=29.7cm,natwidth=21cm
width=7cm,height=9.9cm]{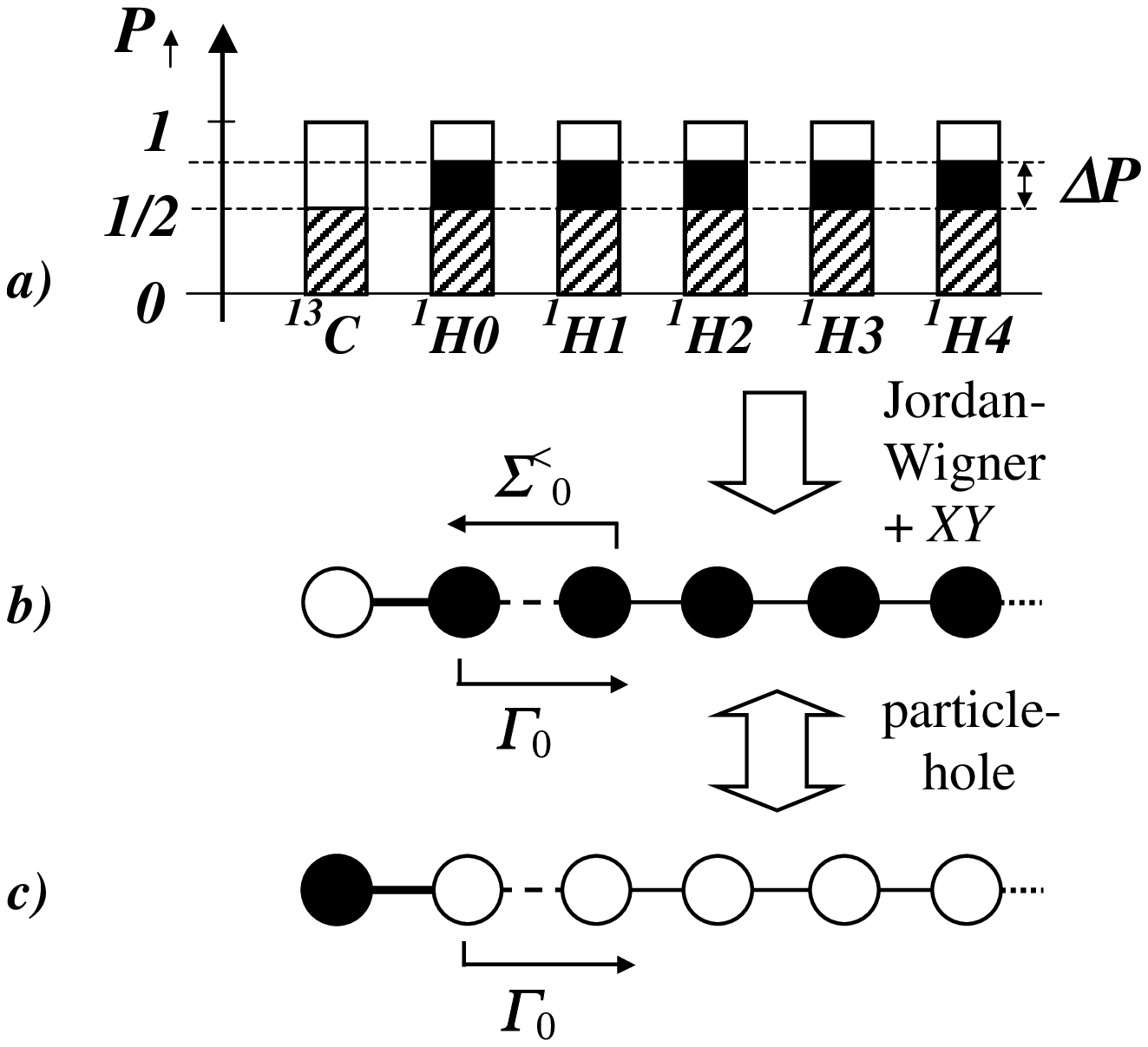}\caption{$a)$ Schematic representation of
the spin system at time $t=0.$ The shaded region stands for the thermodynamic
equilibrium state at high temperature and establish a background probability
level. The black filling represents the excess of probability over the
equilibrium state which is responsible for the observed dynamics. In $b)$ the
same system as in $a)$ after the JWT, that is, under the particle point of
view. Note that in this situation the background contribution is removed and
the dynamics is described by the excess of probability $\Delta P$. In $c)$ we
represent the complementary problem of the case $b)$. Here the black filling
stands for the hole that represent the excitation. In this representation it
is easier to calculate the memory effects in the bath (see text).}%
\label{fig--System}%
\end{figure}

The evaluation of the incoherent contribution requires some explanation about
the model for the $\Sigma_{k,l}^{\;<\;}(\varepsilon,t_{i})$ function.
Following Eq. (3.15) in Ref. \cite{z--GLBEII}
\[
\Sigma_{k,l}^{\;<\;}(\varepsilon,t_{i})=\mathrm{i}2\Gamma_{0}(\varepsilon
)\mathrm{f}_{1}(\varepsilon,t_{i})\delta_{k,0}\delta_{0,l},
\]
with $\Gamma_{0}(\varepsilon)=\Gamma$ as was previously presented and
$\mathrm{f}_{n}(\varepsilon,t_{i})=\left(  \frac{1}{2}+\Delta P\right)
\theta(t_{i})$ stands for the occupation factor of the spin bath. Initially,
all the $^{1}$H nuclei are equally polarized and this represents the initial
condition at $t=0$ for the environment. A schematic representation of this
situation can be seen in Fig. \ref{fig--System} $(a).$ All the dynamics arises
from the excess of probability $\Delta P$ at the $^{1}$H sites. It is
interesting to note, as will be explained later, that the background
probability (shaded region) does not contribute to the dynamics of the
system,\ neither in $P_{0,0}(t)$ nor in $P_{-1,0}(t).$ Having this in mind,
the initial condition, in the particle language can be expressed with a
\textit{normalized }occupation factor $\mathrm{f}_{n}(\varepsilon
,t_{i})=1\times\theta(t_{i})$ for the $0$-th site ($^{1}$H nucleus) and the
spin bath, while the $-1$-th site is empty of excitation $\mathrm{f}%
_{-1}(\varepsilon,t_{i})=0\times\theta(t_{i})$, as can be seen in Fig.
\ref{fig--System} $(b)$. At $t=0$ we allow the environment to interact with
the system and starts injecting probability into the system through the $0$-th
site. Then we have
\begin{equation}
\Sigma_{k,l}^{\;<\;}(\varepsilon,t_{i})=\mathrm{i}\theta(t_{i})2\Gamma
\delta_{k,0}\delta_{0,l}.\label{eq--Sigma}%
\end{equation}
To evaluate Eq. (\ref{eq--N_PCFunc}) we need $G_{0,0}^{<}(t,t)$ from Eq.
(\ref{eq--DanielET}) which is determined by Eq. (\ref{eq--Sigma}). Taking into
account that for a two spin system the sum in eq. (\ref{eq--GmenosTotal}) has
only two terms, $N=1$ and $N=2$, we obtain%

\begin{equation}
P_{0,0}(t)=1-\dfrac{\exp\{-\frac{t\Gamma}{\hbar}\}}{2\cos^{2}(\phi)}%
+\dfrac{\exp\{-\frac{t\Gamma}{\hbar}\}}{2\cos^{2}(\phi)}\cos(\alpha t),
\label{eq--P00}%
\end{equation}
where we have defined $\alpha=\tfrac{2V}{\hbar}\sqrt{\left(  1-(\tfrac{\Gamma
}{2V})^{2}\right)  }$ and $\phi=\arctan\left\{  \left(  \left(  \tfrac
{2V}{\Gamma}\right)  ^{2}-1\right)  ^{-\frac{1}{2}}\right\}  $.

The same calculations for $G_{-1,-1}^{<}(t,t)$ leads to%

\begin{equation}
P_{-1,0}(t)=1-\dfrac{\exp\{-\frac{t\Gamma}{\hbar}\}}{2\cos^{2}(\phi)}%
-\dfrac{\exp\{-\frac{t\Gamma}{\hbar}\}}{2\cos^{2}(\phi)}\cos(\alpha t-2\phi).
\label{eq--Pm10}%
\end{equation}
For both, $P_{0,0}(t)$ and $P_{-1,0}(t),$ the $-1$ term in the rhs of Eq.
(\ref{eq--N_PCFunc}) cancels out with the sum of the coherent evolution of the
first term in Eq. (\ref{eq--initial-density}) and the term corresponding to
the injection in the $N=2$ sub-space . This justifies Fig. \ref{fig--System}
$(b).$

Note that for $\Gamma\rightarrow0$ the above expressions tend to the dynamics
in two isolated sites:
\begin{align}
P_{0,0}^{{}}(t) &  =\left(  \frac{1}{2}+\frac{1}{2}\cos(2Vt/\hbar)\right)
,\label{eq--Po00}\\
P_{-1,0}^{{}}(t) &  =\left(  \frac{1}{2}-\frac{1}{2}\cos(2Vt/\hbar)\right)
=1-\left\vert \cos(Vt/\hbar)\right\vert ^{2}.\label{eq--Po-10}%
\end{align}
It is also interesting to note that the characteristic time for the decay of
the probability, $\tau_{2}=\frac{\hbar}{\Gamma},$ is exactly twice that of a
single site with the same environment. The interpretation of this is that due
to the symmetry adopted ($E_{0}=E_{-1}$) the particle is half of the time on
each site being less affected by the interaction with the spin bath.

\begin{figure}[ptbh]
\centering
\includegraphics[natheight=29.7cm,natwidth=21cm,
height=9.9cm,
width=7cm]{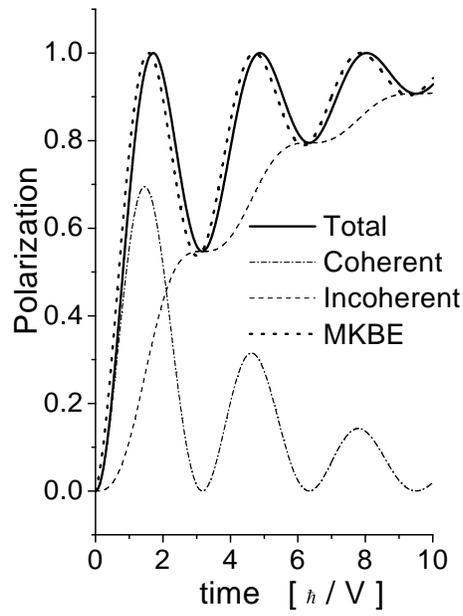}\caption{The $^{13}$C polarization (solid line) is
composed by the sum of a coherent part (dash-dotted line) and an incoherent
contribution (dashed line). For comparison, the evolution obtained with the
secular approximation of\ Ref. \cite{z--MKBE74} (dotted line) is shown denoted
as MKBE. Parameters: $\Gamma/V=0.25$ }%
\label{fig--evolution}%
\end{figure}

Figure \ref{fig--evolution} shows the behavior of the $^{13}$C polarization.
It can be seen that it reaches the value of 1 periodically, converging to the
equilibrium value of 1 at the exponential rate $1/\tau_{2}$. The first maximum
occurs at a relatively short time compared with $\tau_{2}.$ This feature is
used in the spin swap operation by stopping rf irradiation (and hence the
interaction) at a maximal transfer. The maxima in our curves of $P_{-1,0}^{{}%
}(t)$ are always equal to one because of the symmetry adopted ($E_{0}=E_{-1}%
$). However, only the first maxima of the coherent component decaying as
$\exp[-t/\tau_{2}],$ i.e. about 0.7 for our choice of parameters, would be
useful in quantum information processing. The incoherent component of the
polarization, having no definite phase relation with respect to the original
state, bears no information on the quantum evolution. This can be observed by
NMR interferometry as done in Refs. \cite{z--EcosMesoscII,z--LUP-JCP1998}. In
this case the observed polarization at $^{13}$C presents high frequency
oscillations consequence of the interference between the polarization
amplitude that survived at the $^{13}$C and the component returning after
wandering in the $^{1}$H system. This interference would be diminished if, in
the last CP, one uses the second maximum.

\begin{figure}[ptbh]
\centering
\includegraphics[natheight=29.7cm,natwidth=21cm,
height=9.9cm,
width=7cm
]{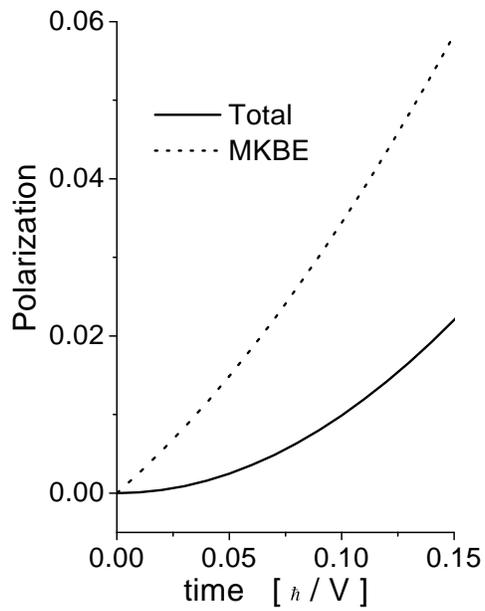}\caption{It is shown the short time regime for the Polarization
function for the $^{13}C$ nucleous obtained within the Keldysh formalism
(solid line) in contrast with the the evolution obtained with the seccular
approximation (MKBE) of Ref. \cite{z--MKBE74} (doted line). Parameters:
$\Gamma/V=0.25$ }%
\label{fig--shorttimes}%
\end{figure}

Another interesting feature of Eqs. (\ref{eq--P00}) and (\ref{eq--Pm10}) is
that they have zero slope for $t=0$ as can be seen in Fig
\ref{fig--shorttimes}. This expresses that the quantum nature of the problem
has not disappeared within the present approximation, in contrast with the
result obtained by using the secular approximation $\Gamma$ $\ll$ $V$,
standard in NMR calculations \cite{z--MKBE74}. Performing the same
approximation as in Ref. \cite{z--MKBE74}, but considering $XY$ coupling
between the system and the spin reservoir, we obtain for the normalized
polarization of the $^{13}$C nucleus
\begin{equation}
P_{-1,0}(t)=1-\frac{\exp\{-\frac{t\Gamma}{\hbar}\}}{2}-\frac{\exp
\{-\frac{t\Gamma}{\hbar}\}}{2}\cos(2Vt/\hbar). \label{eq--MKBE}%
\end{equation}

Both Eqs. (\ref{eq--Pm10}) and (\ref{eq--MKBE}) are obtained considering the
fast fluctuations approximation which leads to $\Gamma_{c}(\varepsilon
)=$constant. However, comparing Eqs. (\ref{eq--Pm10}) and (\ref{eq--MKBE}) it
can be seen that the main differences between them are the \textit{decrease of
the swapping frequency} and the \textit{extra phase} that result from the
Keldysh formalism. The frequency decrease is a natural effect of the damping
in an harmonic oscillator and hence its meaning is clear. The extra phase
provides the correct quadratic behavior for short times. Both effects would
introduce corrections up to a 10\% if one attempts an estimation of the
dipolar frequency (here $2V/\hbar$) from the first experimental maximum.
However, if the frequency is evaluated from the FT of the signal it differs
from the dipolar one in a factor of $(1-(\frac{\Gamma}{2V})^{2})^{1/2}$. This
can have important consequences when one attempts to perform a quantification
of the $^{13}$C$-^{1}$H average distances \cite{z--hirschinger}.

\section{Memory effects of the spin bath}

The $^{13}$C polarization, $P_{-1,0}(t)$, in the Keldysh formalism arises from
the coherent evolution of the initial particle density, for which the
environment is a ``sink'', \ and an incoherent contribution where the bath
acts as a particle ``source''. This can be compared with the complementary
framework. Instead of dealing with a ``particle''\ problem let us consider it
as a ``hole''\ problem (Fig. \ref{fig--System} $(b)$ and $(c)$ respectively).
On these grounds, at $t=0,$ all the sites are occupied except for the
``hole''\ excitation at the $-1$-th site. See Fig. \ref{fig--System} $(c)$
where the black color stands for the hole excitation. At later times this
excitation evolves in the system and also propagates through the reservoir.
The ``environment''\ does not have holes to inject back into the
``system''\ but those evolved coherently from the initial hole (i.e.
$\Sigma^{<}\equiv0$). Here the environment is a perfect ``sink''. Thus all the
dynamics would be coherent, in the sense previously explained. If we add the
result obtained in this case with that of Eq. (\ref{eq--Pm10}) we obtain a
\textit{one} for all times consequence of the particle-hole symmetry. This is
a particularly good test of the consistency of the formalism because in each
result the ``environment''\ is set in a different framework. It also shows
that the background polarization does not contribute to the dynamics.

This \textquotedblleft hole\textquotedblright\ picture can help us to get a
very interesting insight on the dynamics in a case where the memory on the
environment becomes relevant. Consider, for example, the case $V=V_{0}=V_{c}$
and $E_{0}=E_{-1}$. The finite version of this effective Hamiltonian applies
to the actual experiments reported in Ref. \cite{z--Madi-etal}. In this case,
the simplifying approximations of the fast fluctuations regime are not
justified. However, the exact dynamics of the system can be analytically
obtained if one considers an infinite chain. This enables the use of Eq.
(\ref{eq--exact-Self-Energy}) to evaluate the propagator in the first term of
Eq. (\ref{eq--DanielET}). The integration gives the first Bessel function,
hence:
\begin{equation}
P_{-1,0}(t)=1-\left\vert \frac{\hbar}{tV}J_{1}(2tV/\hbar)\right\vert
^{2}.\label{eq--Bessel-1}%
\end{equation}
A first observation is that the frequency above is roughly increased by a
factor of two as compared with that in Eq. (\ref{eq--Po-10}). Since the maxima
of $P_{-1,0}$ are zeroth of the Bessel function it is clear that the frequency
increases slightly with time. These are \textit{memory effects of the
environment }that are dependent on the interplay between the spectral density
of the bath and that of the system.

We notice that the memory effect can also appear in other condition for the
bath. For example, if the proton nuclei have random polarizations and the
density excitation is at site $-1$, i.e. in Fig. \ref{fig--System} ($a$)
\ $\mathrm{f}_{n}(\varepsilon)=\left(  \frac{1}{2}\right)  $ for $n=0,1,..$
representing the $^{1}$H sites filled up to the shaded region; and the $^{13}%
$C site with an occupation $\frac{1}{2}+\Delta P$. In this case the excitation
propagates over a background level (shaded region) that does not contribute to
the dynamics. The schematic view of this initial condition is equivalent to
that of Fig. \ref{fig--System} ($c$) where now the black filling represents a
particle excitation. The solution of the polarization is the first Bessel
function, $P_{-1,-1}(t)=\left\vert \frac{\hbar}{tV}J_{1}(2tV/\hbar)\right\vert
^{2}.$ Apart from the finite size mesoscopic effect, this is precisely the
situation observed\ in Ref. \cite{z--Madi-etal}, although without enough
resolution for a quantitative comparison. The effect of a progressive
modification of the swapping frequency is often observed in many experimental
situations such as CP experiments. Depending on the particular system, the
swapping frequency can accelerate or slow down. Reported examples are Fig. 5
on Ref. \cite{z--LUP-JCP1998} and Fig. 4 on Ref. \cite{z--JCP-2004}. This
simple example solved so far shows that environmental correlations have
fundamental importance in the dynamics and deserve further attention.

\section{Conclusions}

Summarizing, we have solved the Schr\"{o}dinger equation within the Keldysh
formalism with a source boundary condition which results in an injection of
quantum waves without definite phase relation with the initial state. The
model proposed allowed us to consider the effect of the environment over the
system via the decay of the initial state followed with an incoherent
injection. We obtained analytical expressions for the polarization of each of
the components of a $^{13}$C$-^{1}$H system coupled to a spin bath, improving
the result obtained through the application of the secular approximation
\cite{z--MKBE74} in standard density matrix calculation.

Of particular interest is the inclusion of temporal correlations within the
spin bath in a model which has exact solution. On one side, it enabled us to
show a novel result: memory effects can produce a progressive change of the
swapping frequency. On the other side, this results will serve to test
approximate methods developed to deal with complex correlations.

In general, our analytical results based in the spin-particle mapping, allow a
deeper understanding of the polarization dynamics. They may constitute a
starting point for the study of other problems, such as different topologies
\cite{z--Nuestro2002} with XY interaction and the extension to dipolar and
isotropic couplings.

\section*{Acknowledgement}

We acknowledge P. R. Levstein for suggestions on the manuscript. This work
received financial support from CONICET, SeCyT-UNC and ANPCyT.

\bigskip

\section*{Appendix A}

Let us define the function%

\begin{equation}
G_{0,0}^{\;<\,\mathrm{inc.}_{(N)}\,}(t_{2},t_{1})=\sum_{n,m}\int_{0}^{t_{2}%
}\int_{0}^{t_{1}}G_{0,m}^{R_{(N)}}(t_{2},t_{m})\Sigma_{m,n}^{\,\,<_{(N)}%
}(t_{m},t_{n})G_{n,0}^{A_{(N)}}(t_{n},t_{1})\mathrm{d}t_{m}\mathrm{d}%
t_{n},\label{eq--Gmenosinco}%
\end{equation}
which is the second term in Eq. (\ref{eq--Gmenos-Nsubspace}). A similar
expression holds for the coherent part. The manipulation that follows is
independent on the subspace index ($N$), and we will keep it implicit.
Rewriting the integrand in Eq.(\ref{eq--Gmenosinco}) as
\begin{align}
&  G_{0,m}^{R_{{}}}(t_{2},t_{m})\Sigma_{m,n}^{<_{{}}}(t_{m},t_{n}%
)G_{n,0}^{A_{{}}}(t_{n},t_{1})\nonumber\\
&  =G_{0,m}^{R\,\,}(t_{2}-t_{m},\tfrac{t_{2}+t_{m}}{2})\Sigma_{m,n}%
^{\;<\;}(t_{m}-t_{n},\tfrac{t_{m}+t_{n}}{2})G_{n,0}^{A\,}(t_{n}-t_{1}%
,\tfrac{t_{n}+t_{1}}{2})\nonumber\\
&  =\int\int\int G_{0,m}^{R\,\,}(\varepsilon_{R},\tfrac{t_{2}+t_{m}}{2}%
)\Sigma_{m,n}^{\;<\;}(\varepsilon^{\prime},\tfrac{t_{m}+t_{n}}{2}%
)G_{n,0}^{A\,}(\varepsilon_{A},\tfrac{t_{n}+t_{1}}{2})\nonumber\\
&  \exp\left[  -\mathrm{i}\varepsilon_{R}(t_{2}-t_{m})/\hbar\right]
\exp\left[  -\mathrm{i}\varepsilon^{\prime}(t_{m}-t_{n})/\hbar\right]
\exp\left[  -\mathrm{i}\varepsilon_{A}(t_{n}-t_{1})/\hbar\right]
\tfrac{\mathrm{d}\varepsilon_{R}}{2\pi\hbar}\tfrac{\mathrm{d}\varepsilon
^{\prime}}{2\pi\hbar}\tfrac{\mathrm{d}\varepsilon_{A}}{2\pi\hbar
}.\label{eq--integrando}%
\end{align}

Let's define the macroscopic time as $t=\tfrac{1}{2}(t_{2}+t_{1})$ and the
quantum correlation time $\delta t=t_{2}-t_{1}$ which have related time scales
of the injection processes as $t_{i}=\frac{1}{2}\left(  t_{m}+t_{n}\right)  $
and $\delta t_{i}=t_{m}-t_{n}.$ These time scales are associated with
\ $\varepsilon=\frac{1}{2}\left(  \varepsilon_{R}+\varepsilon_{A}\right)  ,$
the energies characterizing the quantum correlation, and $\omega=\frac
{1}{\hbar}\left(  \varepsilon_{R}-\varepsilon_{A}\right)  $ the frequencies in
the observables. The argument in the exponential function becomes
\begin{align*}
&  \varepsilon_{R}t_{2}-\varepsilon_{R}t_{m}+\varepsilon_{A}t_{n}%
-\varepsilon_{A}t_{1}\\
&  =\varepsilon_{R}t+\varepsilon_{R}\tfrac{\delta t}{2}-\varepsilon_{R}%
t_{i}-\varepsilon_{R}\tfrac{\delta t_{i}}{2}+\varepsilon_{A}t_{i}%
-\varepsilon_{A}\tfrac{\delta t_{i}}{2}-\varepsilon_{A}t+\varepsilon_{A}%
\tfrac{\delta t}{2}\\
&  =\hbar\omega(t-t_{i})+\varepsilon\delta t-\varepsilon\delta t_{i},
\end{align*}
and also
\[
\varepsilon^{\prime}(t_{m}-t_{n})=\varepsilon^{\prime}\delta t_{i}.
\]
The Green's functions take the form
\begin{align*}
G_{0,m}^{R\,\,}(\varepsilon_{R},\tfrac{t_{2}+t_{m}}{2}) &  =G_{0,m}%
^{R\,\,}(\varepsilon+\tfrac{\hbar\omega}{2},\tfrac{t+t_{i}}{2}+\tfrac{\delta
t+\delta t_{i}}{4})\\
G_{n,0}^{A\,}(\varepsilon_{A},\tfrac{t_{n}+t_{1}}{2}) &  =G_{n,0}%
^{A\,}(\varepsilon-\tfrac{\hbar\omega}{2},\tfrac{t+t_{i}}{2}-\tfrac{\delta
t+\delta t_{i}}{4})\\
\Sigma_{m,n}^{\;<\;}(\varepsilon^{\prime},\tfrac{t_{m}+t_{n}}{2}) &
=\Sigma_{m,n}^{\;<\;}(\varepsilon^{\prime},t_{i}).
\end{align*}
Finally due to the fact that the transformation have the property that its
Jacobian is equal to one, we have $\mathrm{d}t_{m}\mathrm{d}t_{n}%
=\mathrm{d}t_{i}\mathrm{d}\delta t_{i}$ and $\mathrm{d}\varepsilon
_{R}\mathrm{d}\varepsilon_{A}=\hbar\mathrm{d}\varepsilon\mathrm{d}\omega.$
Replacing all these expressions in the integral of Eq. (\ref{eq--Gmenosinco})
we have
\begin{multline*}
\sum_{n,m}\int_{t_{0}}^{t_{2}}\int_{t_{0}}^{t_{1}}G_{0,m}^{R\,\,}(t_{2}%
,t_{m})\Sigma_{m,n}^{\;<\;}(t_{m},t_{n})G_{n,0}^{A\,}(t_{n},t_{1}%
)\mathrm{d}t_{m}\mathrm{d}t_{n}\\
=\int\int\int\int\int G_{0,m}^{R\,\,}(\varepsilon+\tfrac{\hbar\omega}%
{2},\tfrac{t+t_{i}}{2}+\tfrac{\delta t+\delta t_{i}}{4})\Sigma_{m,n}%
^{\;<\;}(\varepsilon^{\prime},t_{i})G_{n,0}^{A\,}(\varepsilon-\tfrac
{\hbar\omega}{2},\tfrac{t+t_{i}}{2}-\tfrac{\delta t+\delta t_{i}}{4})\\
\times\exp\{-\mathrm{i}[\hbar\omega(t-t_{i})+\varepsilon\delta t-\varepsilon
\delta t_{i}+\varepsilon^{\prime}\delta t_{i}]/\hbar\}\mathrm{d}%
t_{i}\mathrm{d}\delta t_{i}\frac{\mathrm{d}\varepsilon}{2\pi\hbar}%
\frac{\mathrm{d}\varepsilon^{\prime}}{2\pi\hbar}\frac{\mathrm{d}\omega}{2\pi}.
\end{multline*}
Then, we can express Eq. (\ref{eq--Gmenosinco}) as
\[
G_{0,0}^{\;<\,\mathrm{inc.}}(t,\delta t)=\int G_{0,0}^{\;<\,\mathrm{inc.}%
}(\varepsilon,t)\exp\left[  -\mathrm{i}\varepsilon\delta t/\hbar\right]
\frac{\mathrm{d}\varepsilon}{2\pi\hbar},
\]
and using the last two expressions we can identify
\begin{multline*}
G_{0,0}^{\;<\,\mathrm{inc.}\,}(\varepsilon,t)=\\
\int\int\int\int G_{0,m}^{R\,\,}(\varepsilon+\tfrac{\hbar\omega}{2}%
,\tfrac{t+t_{i}}{2}+\tfrac{\delta t+\delta t_{i}}{4})\Sigma_{m,n}%
^{\;<\;}(\varepsilon^{\prime},t_{i})G_{n,0}^{A\,}(\varepsilon-\tfrac
{\hbar\omega}{2},\tfrac{t+t_{i}}{2}-\tfrac{\delta t+\delta t_{i}}{4})\\
\times\exp\left[  -\mathrm{i}[\left(  \hbar\omega(t-t_{i})-\varepsilon\delta
t_{i}+\varepsilon^{\prime}\delta t_{i}\right)  /\hbar\right]  ~\mathrm{d}%
t_{i}\mathrm{d}\delta t_{i}\frac{\mathrm{d}\omega}{2\pi}\frac{\mathrm{d}%
\varepsilon^{\prime}}{2\pi\hbar}.
\end{multline*}
A similar expression holds for the coherent term.

Integrating in energy we obtain $G_{0,0}^{\;<\,\,}(t,t).$ If we consider that
the system Hamiltonian is time independent the last complex expression
simplifies to
\begin{multline*}
G_{0,0}^{\;<\,\mathrm{inc.}\,}(t,t)=\int_{t_{0}}^{t}\int_{-\infty}^{\infty
}\int_{-\infty}^{\infty}\sum_{n,m}G_{0,m}^{R\,\,}(\varepsilon+\tfrac
{\hbar\omega}{2})\Sigma_{m,n}^{\;<\;}(\varepsilon,t_{i})G_{n,0}^{A\,}%
(\varepsilon-\tfrac{\hbar\omega}{2})\\
\exp\left[  -\mathrm{i}\omega(t-t_{i})\right]  \frac{\mathrm{d}\omega}{2\pi
}\frac{\mathrm{d}\varepsilon}{2\pi\hbar}\mathrm{d}t_{i},
\end{multline*}
which is similar to the second term in Eq. (\ref{eq--DanielET}).

\end{document}